\documentclass[12pt]{article}
\usepackage{fullpage}
\usepackage{epsf}
\usepackage{epsfig}
\usepackage{amsthm}
\usepackage{amsfonts}
\usepackage{color}
\usepackage{amsmath}
\usepackage{setspace}
\usepackage{natbib}
\usepackage{hyperref}
\usepackage{algorithmic}
\usepackage[boxed]{algorithm}

\begin{document}

\newcommand{\bm}[1]{\mbox{\boldmath $#1$}}
\newcommand{\mb}[1]{#1}
\newcommand{\bE}[0]{\mathbb{E}}
\newcommand{\bV}[0]{\mathbb{V}\mathrm{ar}}
\newcommand{\bP}[0]{\mathbb{P}}
\newcommand{\ve}[0]{\varepsilon}
\newcommand{\mN}[0]{\mathcal{N}}
\newcommand{\iidsim}[0]{\stackrel{\mathrm{iid}}{\sim}}
\newcommand{\NA}[0]{{\tt NA}}
\newcommand{\cB}{\mathcal{B}}
\newcommand{\R}{\mathbb{R}}

\newcommand{\gnw}[1]{\emph{{\bf GNW:} #1} \\*}

\title{\vspace{-1cm} Modeling an Augmented Lagrangian\\ for 
  Blackbox Constrained Optimization
}
\author{Robert B.~Gramacy\thanks{Corresponding author: The University of
Chicago Booth School of Business, 5807 S.~Woodlawn Ave., Chicago IL, 60605;
{\tt rbgramacy@chicagobooth.edu}}
\and Genetha A.~Gray\thanks{Most of the work was done while at Sandia
  National Laboratories, Livermore, CA}
\and S\'ebastien Le~Digabel\thanks{{GERAD} and D\'epartement de
  math\'ematiques et g\'enie industriel, \'Ecole Polytechnique de
  Montr\'eal, Montr\'eal, QC H3C 3A7, Canada}
\and Herbert K.H.~Lee\thanks{Department of Applied Mathematics and
  Statistics, University of California, Santa Cruz, CA 95064}
\and Pritam Ranjan\thanks{Department of Mathematics and Statistics,
  Acadia University, Wolfville, NS B4P 2R6, Canada}
\and Garth Wells\thanks{Department of Engineering, University of
  Cambridge, Trumpington Street, Cambridge CB2 1PZ, UK}
\and Stefan M.~Wild\thanks{Mathematics and Computer Science Division,
  Argonne National Laboratory, Argonne, IL 60439}}

\maketitle

\begin{abstract}
Constrained blackbox optimization is a difficult problem, with most approaches
coming from the mathematical programming literature.  The statistical
literature is sparse, especially in addressing problems
with nontrivial constraints.  This situation is unfortunate because statistical 
methods
have many attractive properties: global scope, handling noisy objectives,
sensitivity analysis, and so forth. To narrow that gap, we propose a 
combination of
response surface modeling, expected improvement, and the augmented Lagrangian
numerical optimization framework.  This hybrid approach allows 
the statistical model to think globally and the augmented Lagrangian to act
locally.  We focus on problems where the constraints are the primary bottleneck,
requiring expensive simulation to evaluate and substantial modeling effort to
map out.  In that context, our hybridization presents a simple yet effective
solution that allows existing objective-oriented statistical approaches, like
those based on Gaussian process surrogates and expected
improvement heuristics, to be applied to the constrained setting with minor
modification. This work is motivated by a challenging, real-data benchmark
problem from hydrology where, even with a simple linear objective function,
learning a nontrivial valid region complicates the search for a global
minimum.

\end{abstract}

\vspace{1cm}


\section{Introduction}
\label{sec:intro}

The area of mathematical programming has produced efficient algorithms for
nonlinear optimization, most of which have provable convergence properties.
They include algorithms for optimizing under constraints and for handling
so-called {\em blackbox} functions, where evaluation requires running an
opaque computer code revealing little about the functional form of the
objective and/or constraints.  Many modern optimization approaches for
blackbox problems converge without derivative information and require only
weak regularity conditions. Since their search is focused locally, however,
only local solutions are guaranteed.

Statistical approaches to blackbox optimization have the potential to offer
more global scope. Methods based on Gaussian process (GP) surrogates and
expected improvement \citep[EI,][]{jones:schonlau:welch:1998} enjoy global
convergence properties and compare favorably with classical alternatives when
objective evaluations are expensive, simulated by (noisy) Monte Carlo
\citep{picheny:etal:2013} or when there
are many local optima.  In more conventional contexts, however, nonstatistical
approaches are usually preferred. Global search is slower than local search;
hence, for easier problems, the statistical methods underperform.
Additionally, statistical methods are more limited in their ability to handle
constraints. Here, we explore a hybrid approach that pairs a global
statistical perspective with a classical augmented Lagrangian localization
technique for accommodating constraints.

We consider constrained optimization problems of
the form
\begin{equation}
\min_x \left\{
f(x)
:
c(x) \leq 0, x \in \cB
\right\},
\label{eq:ineqprob}
\end{equation}
where $f:\R^d \rightarrow \R$ denotes a scalar-valued objective
function, $c : \R^d \rightarrow \R^m$ denotes a vector\footnote{Vector
inequalities are taken componentwise (i.e., for $a, b \in \R^d$,
$a \leq b$ means $a_i\leq b_i$ for all $i=1, \ldots, d$).}  of
constraint functions, and $\cB \subset \R^d$ denotes a known, bounded,
and convex region. Here we take $\cB = \{x\in\R^d : l \leq x \leq u\}$
to be a hyperrectangle, but it could also include other constraints
known in advance.  Throughout, we will assume that a solution of
(\ref{eq:ineqprob}) exists; in particular, this means that the
feasible region $\{x \in \R^d : c(x)\leq 0\} \cap
\cB$ is nonempty. In~(\ref{eq:ineqprob}), we note the clear distinction made
between the known bound constraints~$\cB$ and the constraints
$c_1(x), \ldots, c_m(x)$, whose functional forms may not be known.

The abstract problem in~\eqref{eq:ineqprob} is challenging when the
constraints $c$ are nonlinear, and even more difficult when evaluation of at
least one of~$f$ and~$c$ requires blackbox simulation. In
Section~\ref{sec:review} we review local search algorithms from the numerical
optimization literature that allow for blackbox~$f$ and~$c$. Until very
recently the statistical literature has, by contrast, only offered solutions tailored
to certain contexts.  For example, \cite{scho:welc:jone:1998} adapted EI for
blackbox~$f$ and known~$c$; \cite{gramacy:lee:2011} considered blackbox~$f$
and blackbox $c \in \{0,1\}$; and \cite{will:sant:notz:lehm:2010} considered
blackbox~$f$ and~$c$ coupled by an integral operator. These methods work well
in their chosen contexts but are limited in scope.

The current state of affairs is unfortunate because statistical methods have
much to offer.  Beyond searching more globally, they can offer robustness,
natively facilitate uncertainty quantification, and enjoy a near monopoly in
the noisy observation case.
In many
real-world optimization problems, handling constraints presents the biggest
challenge; many have a simple, known objective $f$ (e.g., linear, such as
total cost $f(x) = \sum_i x_i$) but multiple complicated, simulation-based
constraints (e.g., indicating if expenditures so-allocated meet
policy/physical requirements). And yet, to our knowledge, this important case
is unexplored in the statistical literature. In Section~\ref{sec:hydro} we
present a hydrology problem meeting that description: despite having a simple
linear objective function, learning a highly nonconvex feasible region
complicates the search for a global minimum.

One way forward is to force the problem~(\ref{eq:ineqprob})
into an existing statistical framework.
For example, one could treat $c(x)$ as binary 
\citep{gramacy:lee:2011}, ignoring
information about the \emph{distance} to the boundary separating
feasible (``valid'') and infeasible (``invalid'') regions.
To more fully utilize all available information, we propose a statistical
approach based on the augmented Lagrangian
\citep[AL, e.g.,][]{Bertsekas82}, a tool from mathematical programming
that converts a problem with general constraints into a sequence of
unconstrained (or simply constrained) problems.

For the unconstrained problem(s) we develop novel surrogate modeling and EI
techniques tailored to the form of the AL, similar to the way that
\citet{parr:etal:2012} deploy EI on a
{\em single} conversion from constrained to unconstrained problems.   Borrowing the AL
setup, and utilizing an appropriate sequence of unconstrained problems,
weakens the burden on the statistical optimizer---deployed in this context as
a subroutine---and leverages convergence guarantees from the mathematical
programming literature. Under specific conditions we can derive closed-form
expressions, like EI, to guide the optimization subproblems, and we explore
numerical/Monte Carlo alternatives for other cases.  Importantly, our use of
Monte Carlo is quite unlike optimization by stochastic search, e.g., simulated
annealing \citep[SA,][]{kirk:1983}. 
SA, and other methods utilizing inhomogeneous
Markov chains, offer global convergence guarantees asymptotically. 
However, in our experience and indeed in
our empirical studies herein, such schemes are less than 
ideal when expensive blackbox evaluation severely limits the number 
of simulations that can be performed.

Although the approach we advocate is general, for specificity in this
paper we focus on blackbox optimization problems for which the objective~$f$
is known while the constraints $c$ require simulation.  This setting all
but rules out statistical comparators that emphasize modeling $f$ and
treat~$c$ as an inconvenience.  Throughout, we note how our approach
can be extended to unknown~$f$ by pairing it with standard surrogate modeling
techniques.

The remainder of the paper is organized as follows.  We first describe a 
synthetic  
problem that introduces the
challenges in this area. Then, in Section~\ref{sec:review}, we review  
statistical
optimization and introduce the AL framework for handling constraints.
Section~\ref{sec:ssapm} contains the bulk of our methodological contribution,
combining statistical surrogates with the AL.  Section~\ref{sec:empirical}
describes implementation details and provides results for our toy example.
Section~\ref{sec:hydro} provides a similar comparison for a challenging real-data
hydrology problem.  We conclude in Section~\ref{sec:discuss} with a discussion
focused on the potential for further extension.

\paragraph{A toy problem.}
\label{sec:swprob}

Consider the following test problem of the form~(\ref{eq:ineqprob}) with a
(known) linear objective in two variables, $f(x) = x_1 + x_2$, bounding box
$\cB = [0,1]^2$, and two (blackbox) nonlinear constraints given by
\[c_1(x) = \frac{3}{2} - x_1 - 2x_2 - \frac{1}{2}\sin\left(2\pi(x_1^2-2x_2)\right),
\quad c_2(x) = x_1^2+x_2^2-\frac{3}{2}.\]
\begin{figure}[tb!]
\centering
\begin{minipage}{5.5cm}
\begin{align*}
 x^A&\approx [ 0.1954, \, 0.4044 ], \\ f\left( x^A \right) &\approx 0.5998, \\
 x^B&\approx [ 0.7197, \, 0.1411 ], \\ f\left( x^B \right) &\approx 0.8609, \\
 x^C &= [ 0, \, 0.75 ], \\ f\left( x^C \right) &= 0.75,
\end{align*}
\vspace{0.25cm}
\end{minipage}
\hspace{1.1cm}
\begin{minipage}{6.5cm}
\includegraphics[scale=0.32, trim=100 170 300 200]{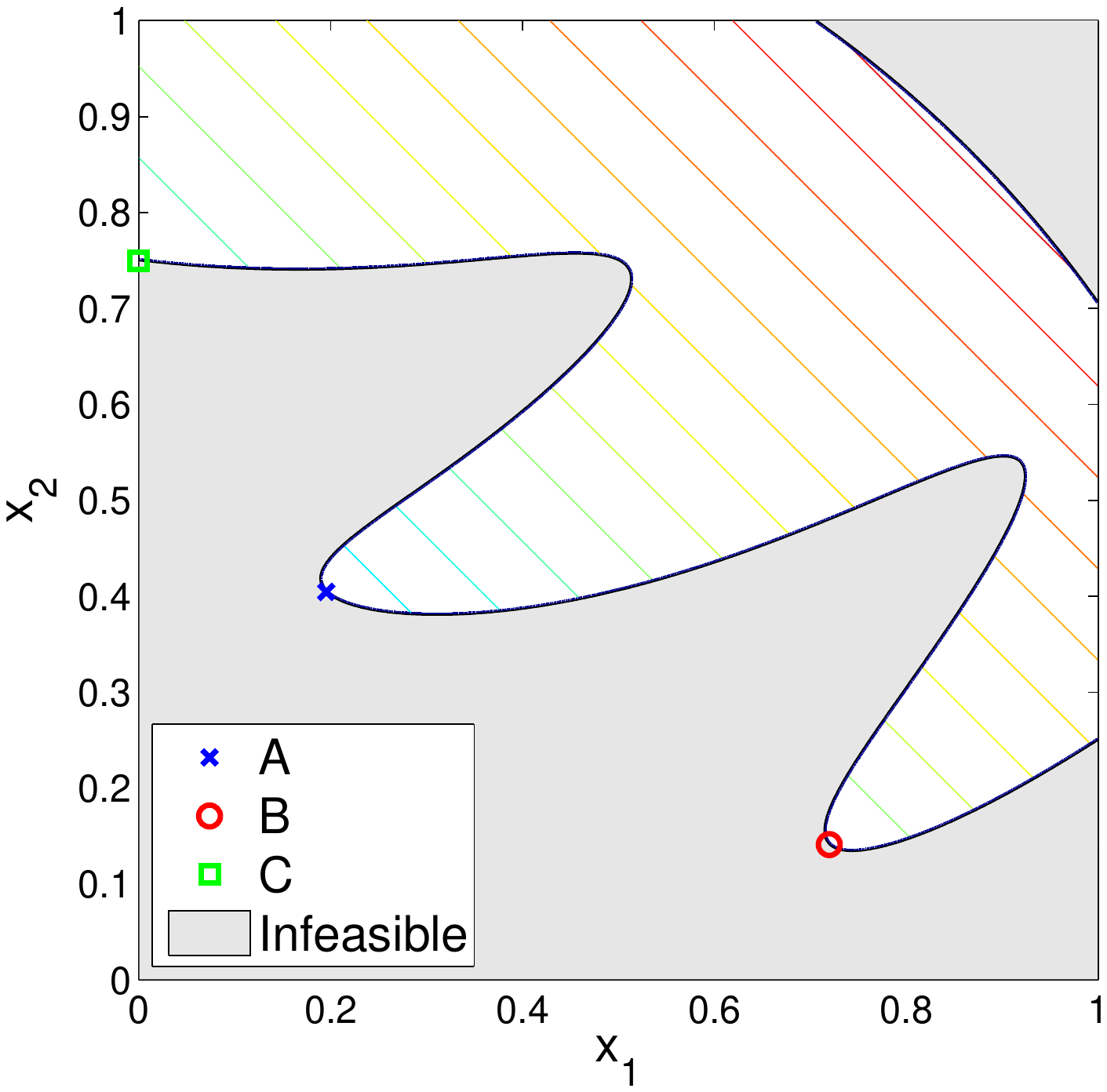}
\end{minipage}
\vspace{-0.25cm}
\caption{Toy problem and its local minimizers;  only $x^A$ is 
a global minimizer.}
\label{f:swprob}
\end{figure}
Figure~\ref{f:swprob} shows the feasible region and the three local optima,
with~$x^A$ being the unique global minimizer. We note that at each of these
solutions, the second constraint is strictly satisfied and the first
constraint holds with equality. For~$x^C$, the lower bound on~$x_1$ is also
binding because if this bound were not present, $x^C$ would not be a local
solution.  The second constraint may seem uninteresting, but it reminds us
that the solution may not be on every constraint boundary and thereby presents
a challenge to methods designed to search that boundary in a blackbox setting.
This toy problem has several characteristics in common with the real-data hydrology
problem detailed in Section~\ref{sec:hydro}. Notably, the two problems both
have a linear objective and highly nonlinear, nonconvex constraint boundaries.

\section{Elements of hybrid optimization}
\label{sec:review}

Here we review the elements we hybridize: response surface models,
expected improvement, and the augmented Lagrangian. Implementations
details are deferred to Section~\ref{sec:empirical}.

\subsection{Surrogate modeling framework for optimization}

Examples of statistical models guiding optimization date back 
at least to \citet{mockus:tiesis:zilinskas:1978}. 
The technique has since evolved, but the basic idea still involves
training a flexible regression model $f^n$ on input-output
pairs $\{\mb{x}^{(i)}, y^{(i)}\}_{i=1}^n$ and using aspects of 
$f^n$ to help choose $x^{(n+1)}$.  One option
is to search the mean of a predictive surface $f^n(x)$ derived from
$f^n$, serving as a {\em surrogate} for the true $f(x)$, for minima.
Gaussian process (GP) regression models provide attractive $f^n$
for a deterministic objective, $f$, since GPs produce highly accurate,
conditionally normal predictive distributions and can interpolate if
desired \citep[see, e.g.,][]{sant:will:notz:2003}. 
This so-called surrogate modeling framework for optimization is focused primarily on the
objective $f$.  Extensions have been made to accommodate constraints
\citep[e.g.,][]{audet:etal:2000}, often by restricting search to the valid region.

\subsection{Expected improvement}
\label{sec:ei}

In initial usage, outlined above, the full statistical potential of $f^n$
remained untapped: estimated uncertainties---a hallmark of any statistical
endeavor---captured in the predictive distributions were not being used.
\citet{jones:schonlau:welch:1998} changed this state of affairs by recognizing 
that the
conditionally normal equations provided by a GP surrogate $f^n(x)$,
completely described by mean function  $\mu^n(x)$ and variance function 
$\sigma^{2n}(x)$, could be used together to balance exploitation and
exploration toward a more efficient global search scheme.
They defined an improvement statistic $I(\mb{x}) = \max\{ 0, f^n_{\min}
- Y(x) \}$, where $f^n_{\min}$ is the minimum among the $n$
$y$-values seen so far, and $Y(\mb{x}) \sim f^n(\mb{x})$ is
a random variable. The improvement assigns large values to inputs
$\mb{x}$, where $Y(\mb{x})$ is likely below $f^n_{\min}$.
\citeauthor{jones:schonlau:welch:1998} showed that the {\em expected
improvement} (EI) could be calculated analytically in the Gaussian case:
\begin{equation}
\bE\{I(x)\} = (f^n_{\min} - \mu^n(x)) \Phi\left(
\frac{f^n_{\min} - \mu^n(x)}{\sigma^n(x)}\right)
+ \sigma^n(x) \phi\left(
\frac{f^n_{\min} - \mu^n(x)}{\sigma^n(x)}\right),
\label{eq:ei}
\end{equation}
where $\Phi$ and $\phi$ are the standard normal cdf and pdf, respectively. The
equation reveals a balance between exploitation ($\mu^n(x)$ under
$f^n_{\min}$) and exploration ($\sigma^n(x)$).  

Leveraging (\ref{eq:ei}), \citeauthor{jones:schonlau:welch:1998}~proposed an 
{\em efficient global optimization} (EGO) scheme
using branch-and-bound to maximize
$\bE\{I(\mb{x})\}$.  In a later paper, \citet{scho:welc:jone:1998} provided an
analytical form for a {\em generalized EI} based on a powered up improvement 
 $I_g(x) =  |f^n_{\min} - Y(x)|^g\mathbb{I}_{\{Y(x) < f^n_{\min}\}}$.  Special
cases of $\bE\{I_g(x)\}$ for $g=0,1,2$ lead to searches via
$\mathrm{Pr}(Y(x) < f^n_{\min})$, the original EI, and the hybrid 
 $\bE\{I(x)\}^2 + \bV[I(x)]$, respectively.

Under weak regularity conditions, search algorithms based on EI
converge to the global optimum. EGO, which specifically pairs GP
surrogates with EI, can be seen as one example of a wider family of 
routines. For example, radial basis function surrogates have been used with
similar success in the context of local search \citep{SMWCAS13}.
Although weak from a technical viewpoint, the computer model regularities
required are rarely reasonable in practice. They ignore potential feedback
loops between surface fits, predictive distributions, improvement
calculations, and search;
in practice, these can pathologically slow
convergence and/or lead to local rather than global solutions.  
Practitioners instead increasingly prefer hybrids between 
global EI and local search
\citep[e.g.,][]{gra:ledigabel:2011}.


\subsection{Augmented Lagrangian framework}
\label{sec:auglag}

{\em Augmented Lagrangian methods} \citep[see, e.g.,][]{nocedal2006no} are a
class of algorithms for constrained nonlinear optimization that enjoy
favorable theoretical properties for finding local solutions from arbitrary
starting points. The main device used by these methods is the augmented
Lagrangian, which, for the inequality constrained problem (\ref{eq:ineqprob}),
is given by
\begin{equation}
L_A(\mb{x};\lambda, \rho) = f(\mb{x}) +\lambda^\top c(\mb{x}) + \frac{1}{2\rho}
\sum_{j=1}^m \max \left(0,c_j(x) \right)^2,
\label{eq:la}
\end{equation}
where $\rho>0$ is a \emph{penalty parameter} and
$\lambda\in\R_+^m$ serves the role of \emph{Lagrange multiplier}.

The first two terms in (\ref{eq:la}) correspond to the Lagrangian,
which is the merit function that defines stationarity for constrained
optimization problems.  Without the second term, (\ref{eq:la}) reduces
to an {\em additive penalty method} (APM) approach to constrained
optimization.
Unless considerable care is taken in choosing the
scale of penalization, however, APMs can introduce ill-conditioning in
the resulting subproblems.

We focus on AL-based methods in which the original nonlinearly
constrained problem is transformed into a sequence of nonlinear
problems where only the bound constraints $\mathcal{B}$ are imposed.
In particular, given the current values for the penalty parameter,
$\rho^{k-1}$, and approximate Lagrange multipliers, $\lambda^{k-1}$,
one approximately solves the subproblem
\begin{equation}
\min_x \left\{ L_A(x;\lambda^{k-1},\rho^{k-1}) : x\in \cB\right\}.
 \label{eq:auglagsp}
 \end{equation}
Given a candidate solution $x^k$, the penalty parameter and approximate
Lagrange multipliers are updated and the process repeats.
Algorithm~\ref{alg:baseal} gives a specific form of these updates. Functions
$f$ and $c$ are evaluated only when solving (\ref{eq:auglagsp}), comprising
the ``inner loop'' [step 2]. 

\begin{algorithm}[ht!]
\begin{algorithmic}[1]
\REQUIRE $\lambda^0\geq 0$, $\rho^0>0$
\FOR{$k=1, 2, \ldots $ (i.e., each ``outer'' iteration)}
\STATE Let $x^k$ (approximately) solve (\ref{eq:auglagsp})
\STATE Set $\lambda_j^k =
\max\left(0,\lambda_j^{k-1}+\frac{1}{\rho^{k-1}} c_j(x^k) \right)$, $j=1,
\ldots, m$
\STATE If $c(x^k)\leq 0$, set $\rho^k=\rho^{k-1}$; otherwise, set
$\rho^k=\frac{1}{2}\rho^{k-1}$
\ENDFOR
\end{algorithmic}
\caption{Basic augmented Lagrangian framework.}
\label{alg:baseal}
\end{algorithm}

We note that termination conditions have not been explicitly provided in
Algorithm~\ref{alg:baseal}. In our setting of blackbox optimization,
termination is dictated primarily by a user's computational budget. Our
empirical comparisons in Sections \ref{sec:empirical}--\ref{sec:hydro} involve
tracking the best (valid) value of the objective over increasing budgets
determined by the number of evaluations of the blackbox (i.e., the cumulative
number of inner iterations). Outside that context, however, one could stop the
outer loop when all constraints are sufficiently satisfied and the
(approximated) gradient of the Lagrangian is sufficiently small; for example,
given thresholds $\eta_1,\eta_2\geq 0$, one could stop when $ \left
\|\max\left\{c(x^k),0\right\}\right\| \leq \eta_1 $ and
$\left\|\nabla f(x^k)+\sum_{j=1}^m \lambda_i^k \nabla c_j(x^k)\right\|\leq
\eta_2.$  Determining convergence within the inner loop [step 2],
is solver dependent; common solvers in our motivating context are discussed
below. The theory for global convergence of the overall AL scheme is forgiving
about the criteria used to end each local search.

\subsection{Derivative-free augmented Lagrangian methods}
\label{sec:dfal}

The inner loop [step 2] of Algorithm~\ref{alg:baseal} can accommodate a host of
methods for solving the simply constrained
subproblem (\ref{eq:auglagsp}).  Solvers can leverage derivatives of
the objective and/or constraint functions when available, or be
\emph{derivative-free} otherwise. We specifically focus on the
derivative-free case because this subsumes blackbox optimization
\citep[see, e.g.,][]{Conn2009a}.  In our comparisons in Sections
\ref{sec:empirical}--\ref{sec:hydro} we consider two benchmark
solvers for the inner loop. We now briefly introduce how these solvers can be
situated within the AL framework; software/implementation details and
convergence detection are provided in the supplementary material,
which also contains the details of three additional comparators
that do not leverage the AL.

{\bf Direct Search:} Loosely, direct search involves probing the
objective at stencils centered on the current best input value.
The outputs obtained on the stencil determine the placement and size
of the next stencil. 
In particular, we consider the mesh
adaptive direct search (MADS) algorithm \citep{AuDe2006}. MADS is a
directional direct-search method that uses dense sets of directions
and generates trial points on a spatial discretization called a mesh.
The most important MADS parameters are the initial and minimal poll
sizes, which define the limits for the {\em poll size parameter},
determining the stencil size, and the {\em maximum mesh index},
which limits poll size reductions after a failed iteration (when a
stencil does not find an improved solution). In the context of
Algorithm~\ref{alg:baseal} it makes sense to allow the initial poll
size parameter to take a software-recommended/default value but to
set the maximum mesh index to $k-1$, prescribing a finer subproblem as
outer iterations progress.

{\bf Model-based:} These are closest in spirit to the statistical
methods we propose. Model-based optimization employs local
approximation models, typically based on local polynomials
\citep[e.g.,][]{Conn2009a} or nonlinear kernels such as radial basis
functions \citep[e.g.,][]{SMWCAS13}, which are related to GPs.
Here we consider the trust-region-based method that was previously
used as an AL inner solver by \citet{kannan:wild:2012}. This method
builds interpolating 
quadratic approximation models to the objective and constraints.
The AL subproblem
(\ref{eq:auglagsp}) is then approximately solved by locally solving a
sequence of quadratics.

\section{Statistical surrogate additive penalty methods}
\label{sec:ssapm}
The methods above are not designed for global
optimization, and it is hard to predict which local minima they 
will ultimately converge to when several minima are present. 
Hybridizing with statistical surrogates offers the potential to improve this 
situation. 
The simplest
approach involves deploying a statistical surrogate directly on the AL
(\ref{eq:la}), but this has drawbacks. To circumvent these, we
consider separately modeling the objective $f$ and each constraint 
 $c_j$. We then pursue options for using the surrogate to solve 
(\ref{eq:auglagsp}), either via the predictive mean or EI,
which has an enlightening closed-form expression in a special case.

\subsection{Surrogate modeling the augmented Lagrangian}
\label{sec:sla}

Consider deploying GP regression-based surrogate modeling of the AL (\ref{eq:la}) in 
order to find $x^k$.  In each iteration of the inner loop (step {\small 2} of 
Algorithm~\ref{alg:baseal}), 
proceed as follows. Let $n$ denote the total number of
blackbox evaluations obtained throughout all previous ``inner'' and
``outer'' iterations, collected as $(x^{(1)}, f^{(1)}, c^{(1)}), \dots,
(x^{(n)}, f^{(n)}, c^{(n)})$. Then form $y^{(i)} = L_A(x^{(i)}; \lambda^{k-1},
\rho^{k-1})$ via $f^{(i)}$ and $c^{(i)}$, and fit a GP surrogate 
to the $n$
pairs $\{(x^{(i)}, y^{(i)})\}_{i=1}^n$.  Optimization can be guided by
minimizing $\mu^n(x)$ in order to find $x^{(n+1)}$ or via EI following
Eq.~(\ref{eq:ei}) with $Y(x)\equiv Y_{\ell^n}(x)\sim\mathcal{N}(\mu^n(x),
\sigma^{2n}(x))$. Approximate convergence can be determined by various 
simple heuristics, from the number of iterations passing without actual improvement,
to monitoring the maximal EI \citep{gramacy:polson:2011} over the trials.

At first glance this represents an attractive option, being modular and
facilitating a global-local tradeoff.  It is modular in the sense that
standard software can be used for surrogate modeling and EI. It is global
because the GP is trained on the entire data seen so far, and EI balances
exploration and exploitation. It is local because, as the AL
``outer'' iterations progress, the (global) ``inner'' searches organically
concentrate near valid regions.

Several drawbacks become apparent, however, upon considering
the nature of the composite objective (\ref{eq:la}). For example, the $y^{(i)}$
values, in their relationship with the $x^{(i)}$s, are likely to exhibit 
behavior that requires nonstationary
surrogate models, primarily because of the final squared term in the
AL, amplifying the effects of $c(x)$ away from the boundary with the valid
region. Most out-of-the-box GP regression methods assume
stationarity, and will therefore be a poor match.
A related challenge is the $\max$ in (\ref{eq:la}),
which produces kinks near the boundary of the valid region, with the regime  
changing behavior across that boundary.

Modern GP methods accommodate nonstationarity
\citep[e.g.,][]{schmidt:ohagan:2003} and even regime-changing
behavior \citep{gra:lee:2008}. To our knowledge, however, only the latter
option is paired with public software.  That method leverages treed
partitioning, whose divide-and-conquer approach can accommodate limited
differentiability and stationarity challenges, but only if regime changes are
roughly axis-aligned. Partitioning, however, does not parsimoniously address
effects amplified quadratically in space.  In fact, no part of the above
scheme, whether surrogate modeling (via GPs or otherwise) or EI-search,
acknowledges the {\em known} quadratic relationship between objective ($f$)
and constraints ($c$).  By treating the entire apparatus as a blackbox, it
discards potentially useful information.  Moreover, when the objective portion
($f$) is completely known, as in our motivating example(s), the fitting method
(unless it accommodates a known mean) needlessly
models a known quantity, which is inefficient \citep[see,
e.g.,][]{kannan:wild:2012}.

\subsection{Separately modeling the pieces of the composite}
\label{sec:sep}

Those shortcomings can be addressed by deploying surrogate models separately on
the components of the AL, rather than wholly to the composite.  With
separate models, stationarity assumptions are less likely to be 
violated since modeling can commence on quantities prior to the problematic
square and $\max$ operations. To clarify, here we take ``separate'' to mean
{\em independent} as that simplifies many matters, although extensions to
correlated models may yield improvements.
Under independence, surrogates $f^n(x)$ for the objective and
$c^n(x) = (c_1^n(x),
\dots, c_m^n(x))$ for the constraints provide distributions
for $Y_{f^n}(\mb{x})$ and $Y_c^n(x) = (Y_{c_1}^n(\mb{x}), \dots,
Y_{c_m}^n(\mb{x}))$, respectively. The $n$ superscripts, which we drop below,
serve here as a reminder that we propose to solve the ``inner'' AL subproblem
(\ref{eq:auglagsp}) using all $n$ data points seen so far.  Samples from those
distributions, obtained trivially via GP surrogates, are easily converted into
samples from the composite, serving as a surrogate for $L_A(x; \lambda, \rho)$:
\begin{equation}
Y(\mb{x}) = Y_f(x) + \lambda^\top Y_c(x) + \frac{1}{2\rho} \sum_{j=1}^m 
\max(0, Y_{c_j}(x))^2.
\label{eq:laem}
\end{equation}
When $f$ is known, we can forgo
calculating $f^n(\mb{x})$ and swap in a deterministic $f(x)$ for
$Y_f(\mb{x})$.

As in Section \ref{sec:sla}, there are several ways to choose new
trial points using the composite {\em distribution} of the random
variable(s) in (\ref{eq:laem}), for example, by searching the predictive mean 
or EI.  We first 
consider the predictive mean approach and defer EI to Section \ref{sec:newei}.  
We have
$\bE \{ Y(\mb{x})\} = \bE\{Y_f(\mb{x})\} + \lambda^\top \bE\{Y_c(\mb{x})\} + 
  \frac{1}{2\rho} \sum_{i=i}^m \bE\{ \max(0, Y_{c_j}(x))^2\}$.
The first two expectations are trivial under normal GP
predictive equations, giving 
\begin{equation}
 \bE \{ Y(\mb{x})\}  = \mu_f^n(\mb{x}) + 
 \lambda^\top \mu_c^n(\mb{x}) + \frac{1}{2\rho} 
  \sum_{j=1}^m \bE\{ \max(0, Y_{c_j}(x))^2\}, \label{eq:aley}
\end{equation}
via a vectorized $\mu_c^n = (\mu_{c_1}^n, \dots, \mu_{c_m}^n)^\top$.
An expression for the final term, which involves $\bE\{ \max(0, Y_{c_j}(x))^2\}$,
can be obtained by recognizing its argument 
as a powered improvement for $-Y_{c_j}(x)$ over zero, that is, 
$I^{(0)}_{-Y_{c_j}}(x) = 
\max\{0, 0 + Y_{c_j}(x)\}$. Since the power is 2, an expectation-variance
relationship can be exploited to obtain  
\begin{align}
\bE\{ \max(0, Y_{c_j}(x))^2\} &= \bE\{I_{-Y_{c_j}}(x)\}^2 + \bV[I_{-Y_{c_j}}(x)] \label{eq:aley2}\\
&= \sigma^{2n}_{c_j}(x)\left[\left(1+ \left(\frac{\mu^n_{c_j}(x)}{\sigma_{c_j}^n(x)}\right)^2 \right)
\Phi\!\left(\frac{\mu^n_{c_j}(x)}{\sigma_{c_j}^n(x)} \right) 
+\frac{\mu^n_{c_j}(x)}{\sigma_{c_j}^n(x)}  
\phi\!\left(\frac{\mu^n_{c_j}(x)}{\sigma_{c_j}^n(x)} \right) \right], \nonumber
\end{align}
by using a result from the generalized EI \citep{scho:welc:jone:1998}.
Combining (\ref{eq:aley}) and (\ref{eq:aley2}) completes the expression for 
$\bE\{Y(x)\}$.
 When $f$ is known, simply replace $\mu_f^n$ with $f$.

\subsection{New expected improvement}
\label{sec:newei}

The composite random variable $Y(x)$ in Eq.~(\ref{eq:laem}) does not have a
form that readily suggests a familiar distribution, for any reasonable choice
of $f^n$ and $c^n$ (e.g., under a GP model), 
complicating analytic calculation of EI. A numerical approximation is
straightforward by Monte Carlo.  Assuming normal predictive
equations, simply sample $y_f^{(t)}(x)$,
$y_{c_1}^{(t)}(x), \dots, y_{c_m}^{(t)}(x)$ from $\mN(\mu_f^n(x),
\sigma_f^{2n}(x))$ and $\mN(\mu_{c_j}^n, \sigma_{c_j}^{2n})$, respectively,
and then average:
\begin{align}
\bE\{I_Y(x)\} &\approx \frac{1}{T} \sum_{t=1}^T \max(0, y_{\min}^n - y^{(t)}(x)) \label{eq:numeric}\\
& = \frac{1}{T} \sum_{t=1}^T 
\max\left[0, y_{\min}^n - \left(y^{(t)}_f(x) + \lambda^\top y^{(t)}_c(x) + \frac{1}{2\rho} \sum_{j=1}^m 
\max(0, y^{(t)}_{c_j}(x))^2\right)\right],  \nonumber
\end{align}
where $y_{\min}^n = \min_{i=1,\dots,n} 
\{ f^{(i)} + \lambda^\top c^{(i)} + \frac{1}{2\rho} \sum_{j=1}^m 
\max(0, c^{(i)}_j)^2\}$ is the best value of the AL observed so far, given the
current $\lambda$ and $\rho$ values. We find generally low Monte Carlo error,
and hence very few samples (e.g., $T=100$) suffice. 

However, challenges arise in exploring the EI surface over $x \in
\mathcal{X}$, since whole swaths of the input space emit numerically zero
$\bE\{I_Y(x)\}$.  When $f$ is known, whereby $Y_f(x) \equiv f(x)$, and when
the outer loop is in later stages (large $k$), yielding 
smaller $\rho^k$, the portion of the input space 
yielding zero EI can
become prohibitively large, complicating searches for improvement. The
quadratic nature of the  AL composite (\ref{eq:laem}) causes $Y$ to be
bounded below for {\em any} $Y_c$-values under certain $(\lambda,
\rho)$, no matter how they are distributed.

To delve a little deeper, consider a single blackbox constraint $c(x)$, a
known objective $f(x)$, and a slight twist on Eq.~(\ref{eq:laem}) where a new
composite $\tilde{Y}$ is defined by removing the $\max$. In this special case,
one can derive an analytical expression for the EI under GP model for $c$.
Let $I_{\tilde{Y}}= \max \{0,\tilde{y}_{\min} - \tilde{Y}\}$ be the
improvement function for the new composite $\tilde{Y}$, suppressing $x$ to streamline
notation. Calculating the EI involves the following integral, where $c(y_c)$
represents the density $c^n$ of $Y_c$, which for a GP is specified by some
$\mu$ and $\sigma^2$:
\[
\bE\{ I_{\tilde{Y}} \} = \int_{-\infty}^\infty I_{\tilde{y}} c(y_c) \; dy_c = 
\int_\theta (\tilde{y}_{\min} - \tilde{y}) \frac{1}{\sqrt{2\pi} \sigma} e^{-\frac{1}{2\sigma^2} (y_c - \mu)^2} \; dy_c,
\quad \theta = \{ y_c: \tilde{y} < \tilde{y}_{\min}\}.
\]
Substitution and integration by parts yield that
when 
$\lambda^2 - 2(f-\tilde{y}_{\min})/\rho \geq 0$,
\begin{align}
\bE\{ I_Y \} &= \left[\tilde{y}_{\min} - \left(\frac{\mu^2}{2\rho} + \lambda\mu + f\right) 
- \frac{\sigma^2}{2\rho}\right](\Phi(v_2) -
\Phi(v_1)) \label{eq:laei2} \\
&\;\;\; + \left[\sigma\mu/\rho + \lambda\sigma](\phi(v_2) - \phi(v_1)\right] 
+ \frac{\sigma^2}{2\rho} ( v_2 \phi(v_2) - v_1 \phi(v_1)), \nonumber \\
\mbox{where} \quad v_1 &= \frac{u_- - \mu}{\sigma}, \quad\quad v_2 = \frac{u_+ - \mu}{\sigma}, \quad\quad
u_\pm = \frac{-\lambda \pm \sqrt{\lambda^2 - 2(f - \tilde{y}_{\min})/\rho}}{\rho^{-1}}. \nonumber
\end{align}
Otherwise, when $\lambda^2 - 2(f-\tilde{y}_{\min})/\rho < 0$, we have that $\bE\{I_{\tilde{Y}}\}
= 0$.  Rearranging gives $\rho\lambda^2 < 2(f - \tilde{y}_{\min})$, which, as $\rho$ is
repeatedly halved, reveals a shrinking region non-zero EI. 

Besides pointing to analytically zero EI values under  (\ref{eq:laem}), the
above discussion suggests two ideas.  First, avoiding $x$ values leading to
$f(x) > y_{\min}$ will boost search efficiency by avoiding zero EI
regions. 
Second, dropping the $\max$ in (\ref{eq:laem})  may lead to
efficiency gains in two ways: from analytic rather than Monte Carlo
evaluation, and via a more targeted search when $f$ is a known monotone
function bounded below over the feasible region.
In that case, a 
solution is known to lie on the boundary
between valid and invalid regions. Dropping the $\max$  will submit large
negative $c(x)$'s to a squared penalty, pushing search away from the interior
of the valid region and towards the boundary.

While the single-constraint, known $f$ formulation is too restrictive for most
problems, some simple remedies are worthy of consideration, especially when
Monte Carlo is not feasible. Extending to blackbox $f$, and modeling $Y_f$, is
straightforward since $Y_f(\mb{x})$ features linearly in Eq.~(\ref{eq:laem}).
Extending to multiple constraints is much more challenging. One option is to
reduce a multiple constraint problem into a single one by estimating a single
surrogate $c^n(\mb{x})$ for an aggregated constraint function, say, $\sum_i
Y_{c_j}(x)$.  Some care is required because summing can result in information
loss which may be extreme if positive and negative $Y_{c_j}(x)$ cancel valid
and invalid values. A better approach would be to use $Y_c = \sum_i
|Y_{c_j}(x)|$ or $Y_c = \sum_i
\max(0, Y_{c_j}(x))$ even though that may result in challenging kinks to
model. The former option, using absolute values, could lead to improvements
when $f$ is a known monotone function, exploiting that the solution is on the
constraint boundary. However, it may introduce complications when the
constraint set includes constraints that are not active/binding at the
solution, which we discuss further in Section~\ref{sec:discuss}. 

\section{Implementation and illustration}
\label{sec:empirical}

In this section, we first provide implementation details for our proposed
methods (Section \ref{sec:ssapm}), and then demonstrate how they fare on our
motivating toy data (Section \ref{sec:intro}).  Implementation details for our
comparators (Section \ref{sec:dfal}) are provided as supplementary material.
which includes details for three further comparators used in our study: a MADS
modification which hands constraints natively, simulated annealing, and an
``asymmetric entropy'' heuristic  \citep{lind:lee:2015}. All methods are
initialized with $\lambda^0 = (0,\dots,0)^\top$ and $\rho^0 = 1/2$.
Throughout, we randomize over the initial $x^0$ by choosing it uniformly in
$\mathcal{B}$.

\subsection{Implementation for surrogate model-based comparators}
\label{sec:simp}

Multiple variations were suggested in Section \ref{sec:ssapm}. We focus our
comparative discussion here on those that performed best.  To be clear, none
of them performed poorly; but several are easily criticized, and those same
methods are consistently dominated by their more sensible counterparts.  In
particular, we do not provide results for the simplistic approach of Section
\ref{sec:sla}, requiring modeling a nonstationary composite AL, since that
method is dominated by separated modeling of the constituent parts, as described
in Section \ref{sec:sep}.

We entertain alternatives from Sections \ref{sec:sep}--\ref{sec:newei} 
that involve guiding the
inner optimization with $\bE\{Y\}$ and $\bE\{I_Y\}$, following 
Eqs.~(\ref{eq:aley2}--\ref{eq:numeric}), respectively.  We note here that the 
results based on a
Monte Carlo $\bE\{Y\}$, via the analog of (\ref{eq:numeric})
without ``$\max[0,y_{\min}^n -$'',
and the analytical alternative
(\ref{eq:aley2}) are indistinguishable up to Monte Carlo error when
randomizing over $x^0$. Taking inspiration from the analytic
EI derivation for the special case in Section \ref{sec:newei}, we entertain
a variation on the numeric EI that discards the $\max$ term.
We do not provide
results based on the analytic expression (\ref{eq:laei2}), however, because 
doing so requires modeling
compromises that hamper effective search.
In total we report results for four variations pairing one of $\bE\{Y\}$
and $\bE\{I_Y\}$ with the original AL (\ref{eq:laem}) and a version obtained
without the $\max$, which are denoted by the acronyms EY, EI, EY-nomax, 
and EI-nomax, respectively.

Throughout we treat $f$ as known and model each $Y_{c_j}$ with separate GPs
initialized with ten random (space filling) input-output pairs from
$\mathcal{B}$ (i.e., the outer loop of Algorithm \ref{alg:baseal} starts with
$x^{(1:10)}$). For fast updates and MLE calculations---when designs are
augmented as Algorithm \ref{alg:baseal} progresses---we used {\tt updateGP}
and {\tt mleGP}, from the {\tt laGP} package \citep{laGP} for {\sf R}.   Each
inner loop search in Algorithm
\ref{alg:baseal} is based on a random set of 1,000 candidate $x$ locations
$\mathcal{X}^n$.  Spacing candidates uniformly in $\cB$ is
inefficient when $f$ is a known linear function.  Instead we consider random
{\em objective improving candidates} (OICs) defined by $\mathcal{X} = \{x :
f(x) < f^{n_*}_{\min}\}$, where $f^{n_*}_{\min}$  is the best value of the
objective for the $n_* \leq n$  {\em valid} points found so far.  If $n_* = 0$, 
then $f^{n_*}_{\min} = \infty$. A random set of candidates $\mathcal{X}^n$ is
easy to populate by rejection sampling even when $\mathcal{X}$ is small
relative to $\cB$. 

A nice feature of OICs is that a fixed number $|\mathcal{X}^n|$ organically
pack more densely as improved $f^{n_*}_{\min}$ are found. However, as progress
slows in later iterations, the density will plateau, with two consequences:
(1) impacting convergence diagnostics based on the candidates (like $\max
\bE\{I_Y\}$) and (2) causing the proportion of $\mathcal{X}^n$ whose EI is
nonzero to dwindle.  We address (1) by declaring approximate convergence,
ending an inner loop search [step 2 of Algorithm \ref{alg:baseal}], if ten
trials pass without improving $y_{\min}^n$.  When guiding search with $\bE\{I_Y\}$, 
earlier approximate convergence [also ending step 2] is declared when
$\max_{x\in \cB}
\bE\{I_Y(x)\} < \epsilon$, for some tolerance $\epsilon$.  Consequence (2) may
be
addressed by increasing $|\mathcal{X}^n|$ over time; however we find it
simpler to default to an $\bE\{Y\}$ search if less than, say, 5\% of
$\mathcal{X}^n$ gives nonzero improvement.  This recognizes that gains to the
exploratory features of EI are biggest early in the search, when the risk of
being trapped in an inferior local mode is greatest.

While the above explains some of the salient details of our implementation,
many specifics have been omitted for space considerations.  For full
transparency please see {\tt optim.auglag} in the {\tt laGP}
package, implementing all variations considered here. \citet{gramacy:2014},
the package vignette, 
provides a worked-code illustration for the toy problem considered below.

\subsection{Empirical results for the toy problem}
\label{sec:toyresults}

\begin{figure}[ht!]
\centering
\begin{minipage}{9.5cm}
\includegraphics[scale=0.86,trim=5 5 0 40]{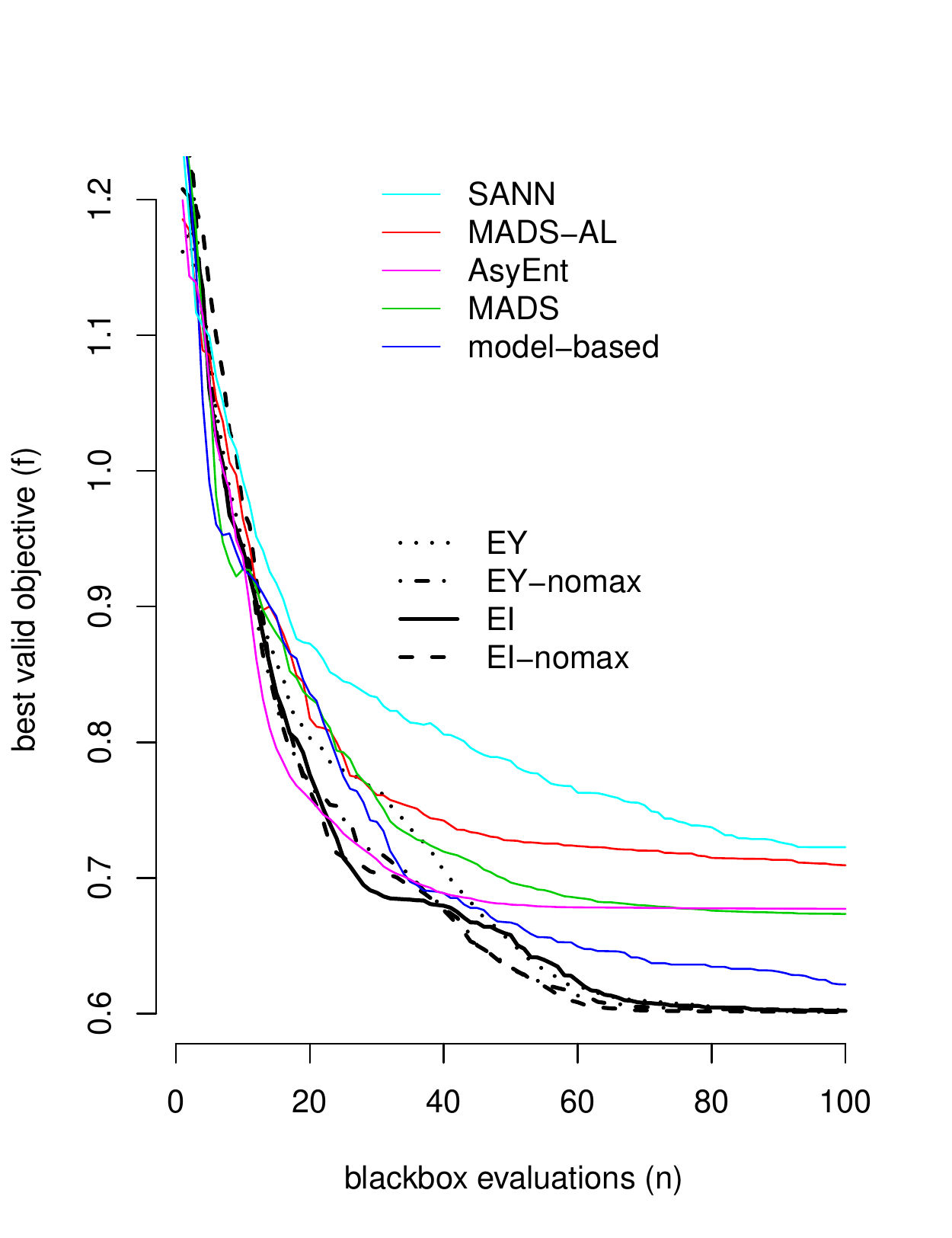}
\end{minipage}
\hfill
\begin{minipage}{5.6cm}
\footnotesize
\begin{tabular}{r|rrr}
  $n$ & 25 & 50 & 100 \\ 
  \hline\hline
  \multicolumn{4}{c}{95\%} \\
  \hline
  EI & 0.866 & 0.775 & 0.602 \\ 
  EI-nomax & 0.906 & 0.770 & 0.603 \\ 
  EY & 1.052 & 0.854 & 0.603 \\ 
  EY-nomax & 1.042 & 0.796 & 0.603 \\ 
  SANN & 1.013 & 0.940 & 0.855 \\ 
  MADS-AL & 1.070 & 0.979 & 0.908 \\ 
  AsyEnt & 0.825 & 0.761 & 0.758 \\
  MADS & 1.056 & 0.886 & 0.863 \\ 
  model & 1.064 & 0.861 & 0.750 \\ 
  \hline\hline
  \multicolumn{4}{c}{average} \\
  \hline
  EI & 0.715 & 0.658 & 0.602 \\ 
  EI-nomax & 0.715 & 0.633 & 0.601 \\ 
  EY & 0.779 & 0.653 & 0.601 \\ 
  EY-nomax & 0.743 & 0.634 & 0.603 \\ 
  SANN & 0.837 & 0.771 & 0.716 \\ 
  MADS-AL & 0.789 & 0.728 & 0.709 \\ 
  AsyEnt & 0.733 & 0.680 & 0.677 \\
  MADS & 0.793 & 0.697 & 0.673 \\ 
  model & 0.775 & 0.667 & 0.621 \\ 
  \hline\hline
  \multicolumn{4}{c}{5\%} \\
  \hline
  EI & 0.610 & 0.602 & 0.600 \\ 
  EI-nomax & 0.613 & 0.601 & 0.600 \\ 
  EY & 0.607 & 0.601 & 0.600 \\ 
  EY-nomax & 0.606 & 0.600 & 0.600 \\ 
  SANN & 0.648 & 0.630 & 0.612 \\
  MADS-AL & 0.600 & 0.600 & 0.600 \\ 
  AsyEnt & 0.610 & 0.601 & 0.600 \\
  MADS & 0.608 & 0.600 & 0.599 \\ 
  model & 0.600 & 0.599 & 0.599 \\ 
   \hline
\end{tabular}
\vfill
\end{minipage}
\caption{Results for the motivating problem in Section \ref{sec:swprob} 
over 100 Monte Carlo repetitions with a random $x^0$.  The plot tracks the
average best valid value of the objective over 100 blackbox iterations; the
table shows distributional information 
at iterations 25, 50, and 100.}
\label{f:swresults}
\end{figure}

Figure \ref{f:swresults} summarizes the results of a Monte Carlo experiment
for the toy problem described in Section \ref{sec:swprob}.
Each of 100 repetitions is initialized randomly in $\mathcal{B} = [0,1]^2$.
The graph in the figure records the average of the best valid value of the
objective over the iterations.  The plotted data coincide with the numbers
shown in the middle section of the accompanying table for the
25$^\mathrm{th}$, 50$^\mathrm{th}$ and final iteration.  The other two
sections show the 90\% quantiles to give an indication of worst- and best-case
behavior.

Figure~\ref{f:swresults} indicates that all variations on our methods
eventually outperform all comparators in terms of both average and worst-case
behavior. All methods find the right global minima in five or more cases (5\%
results), but only the EI-based ones perform substantially better in the worst
case (using the 95\% numbers).  In only one case out of 100 did EI not find
the global minima, whereas 15\% of the model-based runs failed to find it
(these runs finding other local solutions instead).  Except for a brief time
near iteration $n=50$, and ignoring the first 20 iterations for which all
methods perform similarly, EI-based comparators dominate EY analogues. There
is a period ($n\approx 35$) where EI's average progress stalls temporarily. We
observed that this usually marks a transitional period from primarily
exploratory to primarily exploitive behavior. Toward the end of the trials,
the methods based on dropping the $\max$ from Eq.~(\ref{eq:laem}) win out.
Ignoring regions of the space giving large negative values of $c(x)$ seems to
help in these latter stages, but it can hinder performance earlier on.

SA fares worst in this example; however, it does better in our real-data one
below.  Although we summarize results for the first 100 evaluations, we ran
the SA comparator to convergence and report here that each repetition did
eventually converge to the global minimum.  Reaching convergence, however,
took an average of 8,240 evaluations.  Compared with our EI-based results,
this represents an enormous expense, which foreshadows further discussion in
Section
\ref{sec:discuss} about why stochastic search may be less than ideal for
optimization of expensive computer simulation experiments.  The asymmetric entropy
method (``AsyEnt'') performs well
once active search begins after ten space-filling evaluations.
Subsequently its progress plateaus
indicating a struggle to consistently hone-in on solutions near the boundary.

\section{Pump-and-treat hydrology problem}
\label{sec:hydro}
\begin{figure}[ht!]  
\centering
\includegraphics[scale=0.35,trim=0 160 10 210]{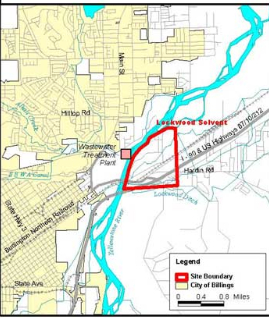}
\hspace{0.75cm}
\includegraphics[scale=0.35,trim=10 160 0 160]{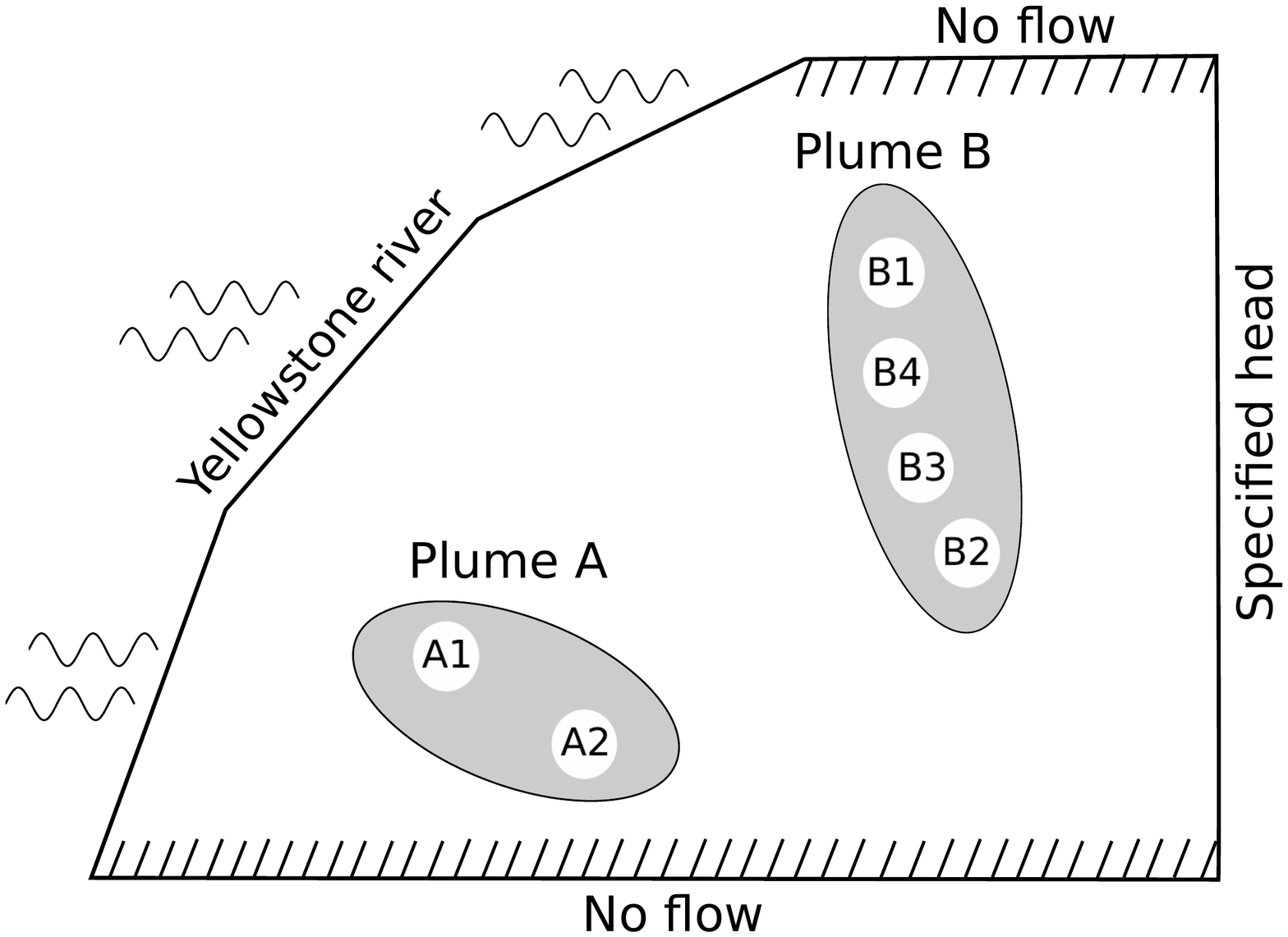}
\caption{Lockwood site and its 
contaminant plumes.  The map on the {\em left} identifies the Lockwood Solvent
region 
and shows its proximity to the Yellowstone 
River and the city of Billings (image from the website 
of \citet{siteMap}). The {\em right} panel illustrates the plume sites,
its boundaries (including the Yellowstone river), and the proposed location 
of six remediation wells (A1, A2, B1, B2, B3, B4).} 
\label{f:lockwood}
\end{figure}

We turn now to our motivating example.  
Worldwide, there are more than 10,000 contaminated land sites \citep{Meer2008}. 
Environmental cleanup at these sites has received increased 
attention over the past 20--30 years.  Preventing the migration of contaminant 
plumes is vital to protect water supplies and prevent disease.  One approach  
is pump-and-treat remediation, in which wells are strategically placed to 
pump out contaminated water, purify it, and inject the treated water back into 
the 
system to prevent contaminant spread.  
A case study of one such remediation is the 580-acre Lockwood 
Solvent Groundwater Plume Site, an EPA Superfund site located near Billings, Montana,
where industrial practices have led to groundwater contamination
 \citep{EPA}. Figure \ref{f:lockwood} shows the location of the site and
provides a simplified illustration of the contaminant plumes that threaten
the Yellowstone River. To prevent further expansion of these plumes,
the placement of six pump-and-treat wells has been proposed, as shown in the
figure.
  
\citet{Mayer2002} posed the pump-and-treat problem as a 
constrained blackbox optimization problem.
For the version of the problem considered here, the pumping rates are varied
in order to minimize the cost of operating the system subject to constraints
on the contaminant staying within the plume boundaries. Letting $x_j$ denote
the pumping rate for well $j$, one obtains the constrained problem
\[\min_{x} \; 
\{f(x) = \sum_{j=1}^6 x_j :  c_1(x)\leq 0, \, c_2(x)\leq 0, \, 
 x\in [0, 2\cdot 10^4]^6 \}.
\]
 The objective $f$ is linear and describes the costs required to operate the
wells.  In the absence of the constraints $c$, the solution is at the origin
and corresponds to no pumping and no remediation. The two constraints denote
flow exiting the two contaminant plumes. An {\em analytic element method}
groundwater model simulates the amount of contaminant exiting the boundaries
and is treated as a blackbox \citep{AEM}. This model never returns negative
values for the constraint, and this nonsmoothness---right at the constraint
boundary---can present modeling challenges.

\subsection{Some comparators}
 \label{sec:lockcompare} 

\begin{figure}[t!]
  \includegraphics[scale=0.35,trim=9 0 0 80]{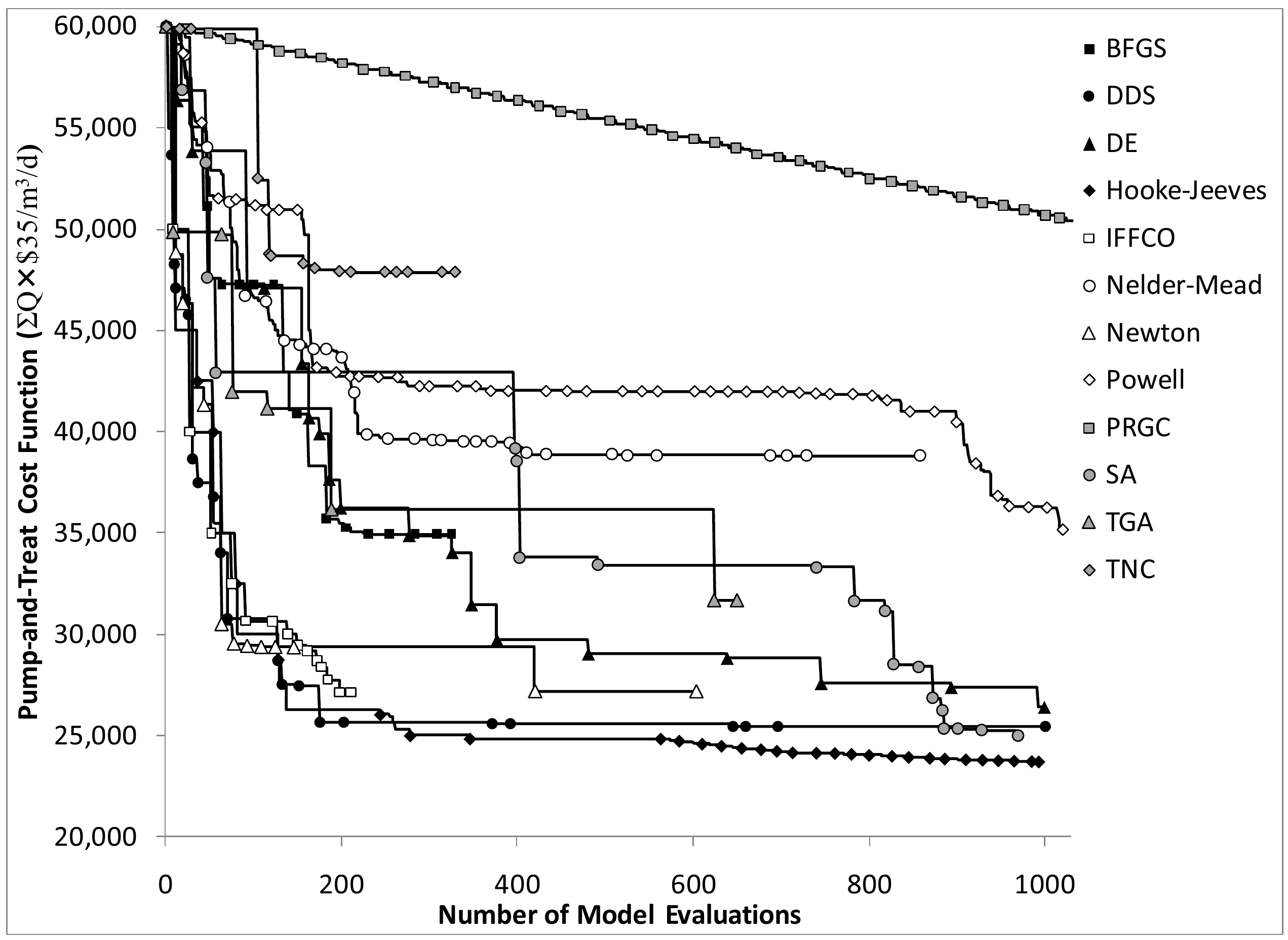}
\hfill
\includegraphics[scale=0.6,trim=5 60 30 40]{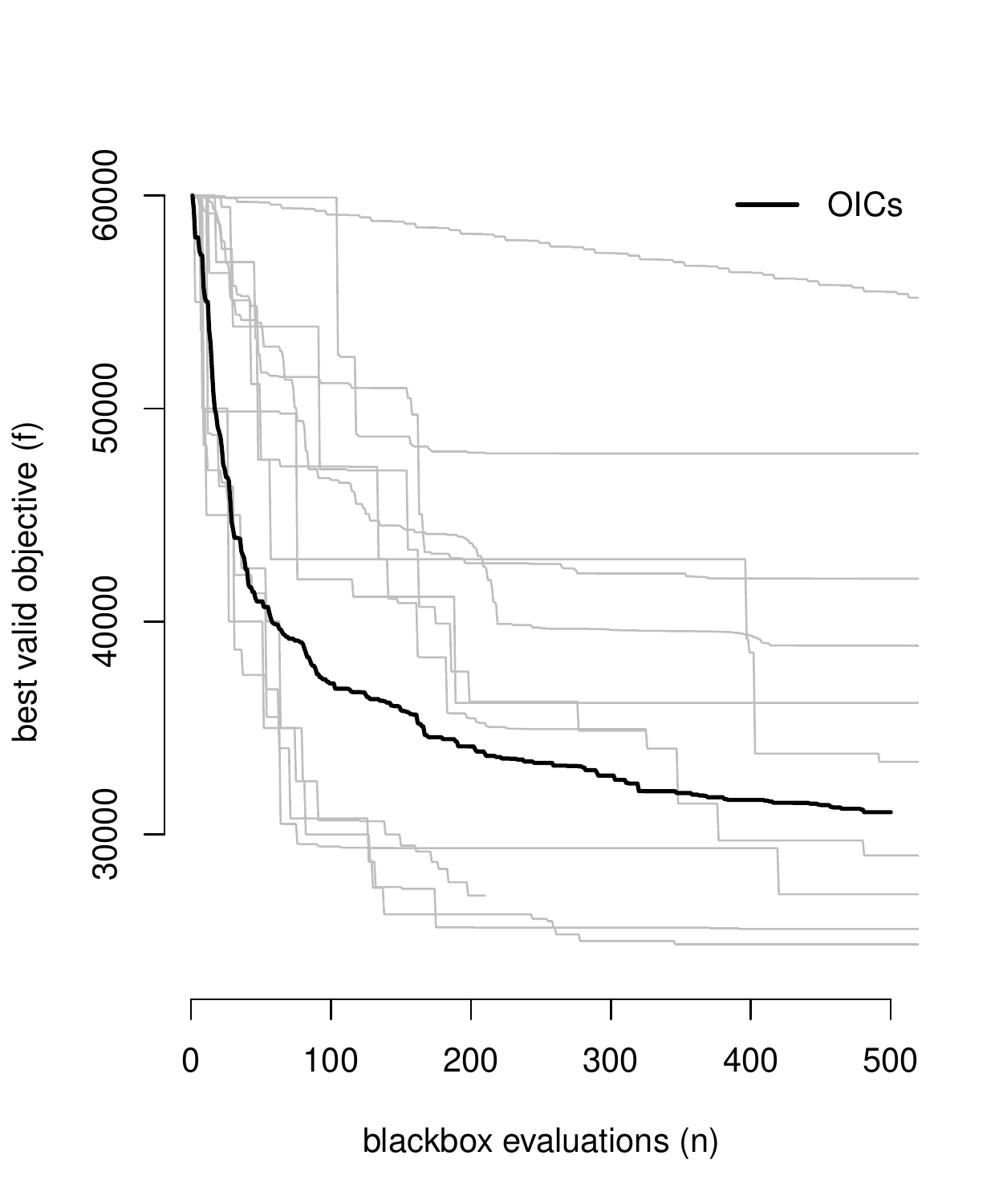}
\vspace{0.7cm}
\caption{Progress of algorithms on the Lockwood problem; the vertical axes 
denote the value of the objective at the best valid iterate as a function of 
the number of optimization iterations. 
The {\em left} graph shows the results of algorithms compared by
\citep{matott:leung:sim:2011}; the {\em right} one abstracts the {\em
 left} graph for further comparison (e.g., using OICs).} 
\label{f:matott}
\end{figure}

\citet{matott:leung:sim:2011} featured this example in a comparison of {\sf
MATLAB} and {\sf Python} optimizers, treating constraints via APM. The results
of this study are shown in the {\em left} panel of Figure~\ref{f:matott}
under a total budget of 1,000 evaluations. All comparators were
initialized at the valid input $x^0 = (10,000, \dots, 10,000)^\top$. 
To abstract these trajectories as a benchmark we shall superimpose our new results on 
an image of the first 500 iterations.

The {\em right} panel of Figure~\ref{f:matott} shows an example, with a simple 
comparator overlaid
based on stochastic search with OICs (Section~\ref{sec:simp}).
It may be surprising that simple stochastic search---based on 
sampling one OIC
in each trial and updating the best valid value of the objective when a new
one is found---performs well relative to much more
thoughtful comparators. Since the method is stochastic, we are 
revealing its average
behavior over thirty replicates. That average is competitive with the best
group of methods for the first twenty-five iterations or so, suggesting that
those methods, while implementing highly varied protocols, are not searching
any better than randomly in early stages. Observe that even after those
early stages, OICs still outperform at least half the comparators for the
remaining trials. Those methods are getting stuck in local minima,
whereas OICs are shy of clumsily global. However pathologically 
slow a random
search like this may be to converge, its success on this problem illustrates a 
clear benefit to exploration over exploitation in early stages.

\subsection{Using augmented Lagrangians}

\begin{figure}[ht!]
\centering
\begin{minipage}{9.4cm}
\includegraphics[scale=0.86,trim=5 5 0 40]{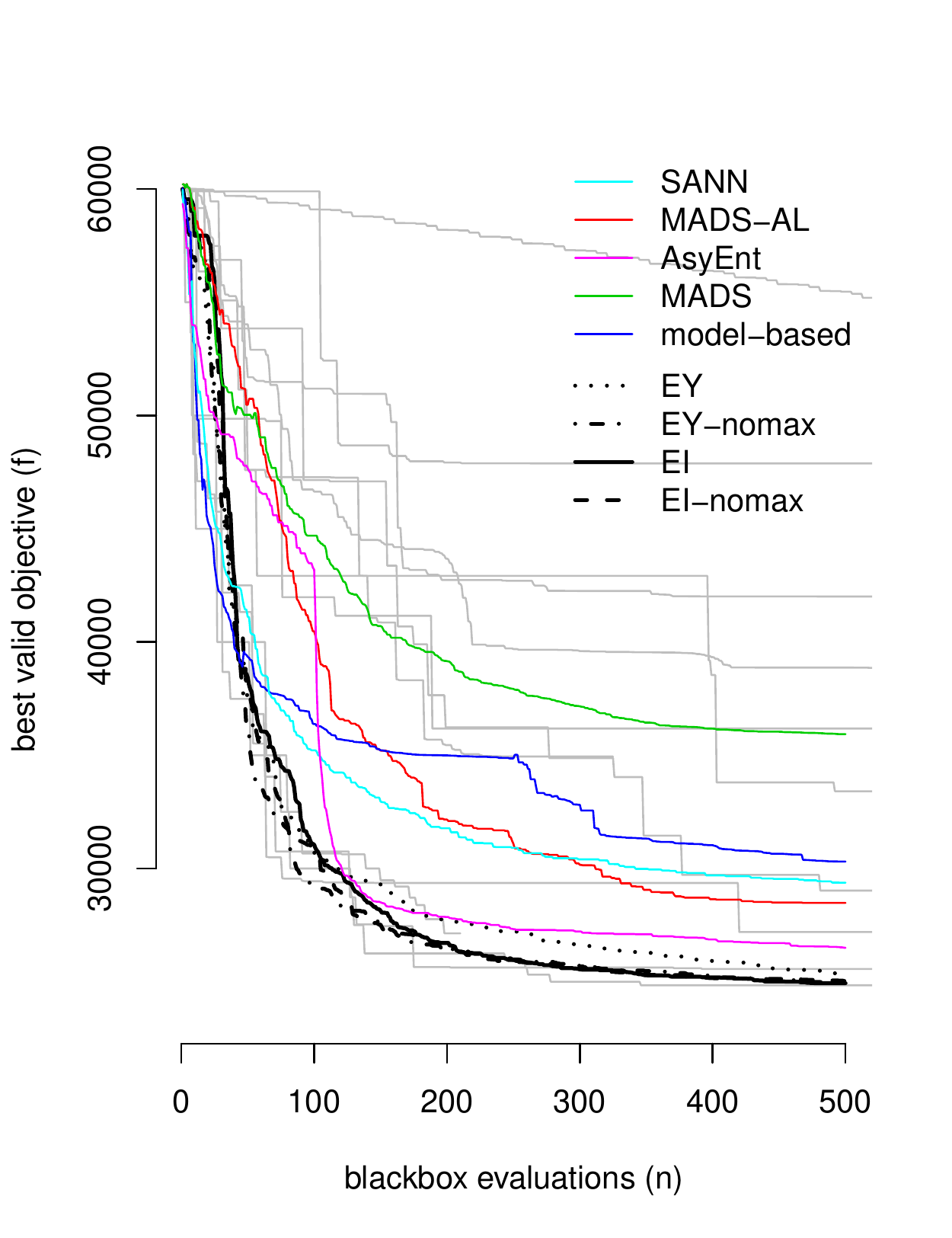}
\end{minipage}
\hfill
\begin{minipage}{5.8cm}
\footnotesize
 \begin{tabular}{r|rrr}
   $n$ & 100 & 200 & 500 \\ 
   \hline\hline
   \multicolumn{4}{c}{95\%} \\
   \hline
EI & 37584 & 28698 & 25695 \\ 
EI-nomax & 43267 & 28875 & 25909 \\ 
EY & 36554 & 32770 & 27362 \\ 
EY-nomax & 36446 & 29690 & 26220 \\ 
SANN & 43928 & 35456 & 30920 \\
MADS-AL & 60000 & 49020 & 32663 \\ 
AsyEnt & 49030 & 29079 & 27445 \\ 
MADS & 60000 & 60000 & 60000 \\ 
model & 60000 & 60000 & 35730 \\ 
   \hline\hline
   \multicolumn{4}{c}{average} \\
   \hline
EI & 31048 & 26714 & 24936 \\ 
EI-nomax & 30874 & 26333 & 25032 \\ 
EY & 30701 & 27721 & 25359 \\ 
EY-nomax & 29199 & 26493 & 24954 \\ 
SANN & 35219 & 31777 & 29375 \\
MADS-AL & 40474 & 32162 & 28479 \\ 
AsyEnt & 43194 & 27860 & 26500 \\ 
MADS & 44694 & 39157 & 35931 \\ 
model & 36378 & 34994 & 30304 \\ 
 \hline\hline
 \multicolumn{4}{c}{5\%} \\
 \hline
EI & 27054 & 25119 & 24196 \\ 
EI-nomax & 26506 & 24367 & 24226 \\ 
EY & 25677 & 24492 & 24100 \\ 
EY-nomax & 25185 & 24487 & 24128 \\ 
SANN & 28766 & 27238 & 26824 \\
MADS-AL & 30776 & 26358 & 24102 \\ 
AsyEnt & 37483 & 26681 & 25377 \\ 
MADS & 30023 & 26591 & 23571 \\ 
model & 25912 & 25164 & 24939 \\ 
    \hline
 \end{tabular}
\vfill
\end{minipage}
\caption{Results for the Lockwood problem over
30 Monte Carlo repetitions with a random $x^0$.  The plot tracks the
average best valid value of the objective over blackbox iterations;
the table shows more distributional information 
at iterations 100, 200, and 500.}
\label{f:lockresults}
\end{figure}

Figure \ref{f:lockresults} shows the results of a Monte Carlo experiment set
up like the one in Section~\ref{sec:toyresults}.  In this case each of
thirty repetitions was initialized randomly with $x^0 \in \mathcal{B} =
[0, 20000]^6$.  
Note that comparators from Section~\ref{sec:lockcompare} (in gray)
used a single fixed starting location $x^0$, {\em not} a random one.
From the figure we can see that the relative ordering of our proposed hybrid
surrogate/AL comparators is roughly the same as for the toy problem.  The
surrogate-model-based average and worst-case behaviors are better than those
of the other AL comparators and are competitive with those of the best APMs
from
\citeauthor{matott:leung:sim:2011} 
Many of the individual Monte Carlo runs of our EI- and EY-based
methods outperformed all APM comparators. 

But some individual APM runs outperform our average results. 
We believe that the initializing value $x^0$ used by
those methods is a tremendous help.  For example, when running MADS  
(no AL) with that same value, it achieved the best result in our study, 23,026. 
That MADS' average behavior is much
worse suggests extreme differential performance depending on the quality of
initialization, particularly with regard to the validity of the initial
value $x^0$.  Moreover, the 95\% section of the table reveals that a
substantial proportion (5-10\%) of the repetitions resulted in no valid
solution even after exhausting the full budget of $n=500$
iterations.\footnote{We put 60,000 in as a placeholder for these cases.} Had the 
\citeauthor{matott:leung:sim:2011}~comparison been
randomly initialized, we expect that the best comparators would similarly have 
fared 
worse.  By contrast, in experiments with the surrogate-based methods using the
same valid $x^0$ we found (not shown) no differences, up to Monte Carlo
error, in the final solutions we obtained.

The SA comparator performed rather better in this experiment compared our
synthetic example in Section \ref{sec:toyresults}, out-performing all but one
of the classical AL methods.   Due to the expense of the blackbox simulations
(and the likelihood that convergence could require thousands of iterations) we
did not attempt to run SA to convergence to see if it is ultimately
competitive with our EI-based methods.  The story is similar for ``AsyEnt''.
After the one hundred space-filling candidates, progress is rapid---beating out
our classical comparators---but it quickly plateaus, never quite matching our
surrogate-model-based AL ones.   Interestingly, SA and AsyEnt have the worst
best-case (5\% and 500 iterations) results.

\section{Discussion}
\label{sec:discuss}

We explored a hybridization of statistical 
global optimization with an amenable mathematical programming
approach to accommodating constraints.  In particular, we combined Gaussian
process surrogate modeling and expected improvement methods from
the design of computer experiments literature with an additive penalty method
that has attractive convergence properties: the augmented Lagrangian.
The main advantage of this pairing is that it reduces a constrained
optimization into an unconstrained one, for which statistical methods are
more mature.  Statistical methods are not known for their rapid
convergence to local optima, but they are more conservative than their
mathematical programming analogues: in many cases offering better global
solutions with fewer blackbox evaluations.

We extended the idea of EI to a composite objective arising from the AL and
showed that the most sensible variations on such schemes consistently
outperform similar methods leveraging a more traditional optimization framework
whose focus is usually more local.  Still, there is plenty of room 
for improvement.  For example, we anticipate gains from
a more aggressive hybridization that acknowledges that the statistical
methods fail to ``penetrate'' into local troughs, particularly toward 
the end of a search.  In the unconstrained context, 
\citet{gray:mart:tadd:lee:gram:2007} and \citet{gra:ledigabel:2011} 
have had success pairing EI with modern direct search optimizers.
Both setups port provable local convergence from the direct method 
to a more global search context by, in effect, letting the direct solver
take over toward the end of the search in order to ``drill down'' to a 
final solution.

Other potential extensions involve improvements on the statistical
modeling front.  For example, our models for the constraints were 
independent for each $c_j$, $j=1,\dots,m$, 
leaving untapped potential to leverage cross correlations 
\citep[e.g.,][]{will:sant:notz:lehm:2010}.  Ideas from multiobjective
optimization may prove helpful in our multiconstraint format.  Treating them
as we do in a quadratic composite (via the AL) represents one way forward;
keeping them separated with Pareto-optimal-like strategies may be
another promising option
\citep[see, e.g.,][]{svenson:santner:2012,picheny:2013,picheny:2014}.

Better analytics offer the potential for further improvement.  We resorted to
Monte Carlo (MC) for two aspects of our calculations, however it is important
to clarify we are not advocating a stochastic search.  An analytic EI
calculation, or a branch-and-bound-like scheme for optimizing it at each
inner-loop search step, would eliminate MC errors and yield an entirely
deterministic scheme.  Sometimes a bit of stochasticity is welcome in
statistical design applications, of which blackbox optimization is an example.
However, there are clearly limits to the usefulness of random
search in that setting, especially when simulations for the
objective or constraint(s) are expensive.  This is borne out in our empirical
work where a simulated annealing (SA) implementation demonstrates lukewarm
results under tight computational budgets. SA offers nice technical guarantees
for global solutions but, as the {\sf R} documentation for {\tt
optim}'s \verb!method="SANN"! cautions, it ``depends critically on the
settings of the control parameters. It is not a general-purpose method but can
be very useful in getting to a good value on a very rough surface.''

There may be alternative ways to acknowledge---in the known monotone objective
$(f)$ case, as in both of our examples---that the solution lies on a
constraint boundary.  Our ideas for this case (e.g., dropping the $\max$ in
the AL (\ref{eq:laem})) are attractive because they can be facilitated by a
minor coding change, but they yield just modest improvements. It is also risky 
when
the problem includes nonbinding constraints at the solution, by
inappropriately inflating the importance of candidates well inside the
valid region according to one constraint, but well outside for another. The
slack variable approach of
\citet{kannan:wild:2012} may present an attractive remedy,
especially when $c(x)$ returns only nonnegative
values like in our hydrology example, as might
explicit learning of classification boundaries to guide sampling 
\citep[e.g.,][]{lee:gram:link:gray:2011,chev:etal:2014,lind:lee:2015}.

In closing, however, we remark that perhaps extra complication, which is what
many of the above ideas entail, may not be pragmatic from an engineering
perspective.  
The AL is a simple framework and its hybridization with GP
models and EI is relatively straightforward, allowing existing statistical
software to be leveraged directly (e.g., {\tt laGP} was easy to augment to
accommodate the new methods described here).  This is attractive
because, relative to the mathematical programming literature, statistical
optimization has few constrained optimization methods
readily deployable by practitioners.  The statistical optimization literature
is still in its infancy in the sense that bespoke implementation is required
for most novel applications. By contrast, software packages like {\tt 
NOMAD}, implementing MADS (see supplementary material)
generally work out-of-the-box.  It is hard to imagine
matching that engineering capability for difficult constrained optimization
problems with statistical methodology if we insist on those methods being even more
intricate than the current state-of-the-art.

\subsection*{Acknowledgments}

We would like to thank the American Institute of Mathematics for hosting us in a
series of small group meetings. Many thanks to an Associate Editor and Referee
for thoughtful comments throughout the review process.  Lee was supported by
National Science Foundation grant DMS-0906720. Wild was supported by the
Applied Mathematics activity within the U.S.\ Department of Energy, Office of
Science, Advanced Scientific Computing Research, under Contract No.\
DE-AC02-06CH11357. Le~Digabel and Ranjan's research is supported by discovery
grants from the Natural Sciences and Engineering Research Council of Canada.
Thanks to David Lindberg for the Lockwood diagram in the right panel of
Figure~\ref{f:lockwood}.

\pagebreak
\appendix
\renewcommand{\thesection}{SM \arabic{section}}
\renewcommand{\thesubsection}{SM\S\arabic{subsection}}
\setcounter{figure}{0}
\setcounter{table}{0}
\setcounter{equation}{0}
\renewcommand{\thefigure}{SM.\arabic{figure}}    
\renewcommand{\thetable}{SM.\arabic{table}}    
\renewcommand{\theequation}{SM.\arabic{equation}}    
\setcounter{page}{1}
\renewcommand{\thepage}{SM \arabic{page}}

\section*{Supplementary Materials}

\subsection{Implementation details for comparators}
\label{sec:imp}

\subsubsection*{Classical AL comparators}

{\bf Direct:} For MADS we use the implementation in the
{\tt NOMAD} software~\citep{Le09b,AuCo04a}. Beyond adaptations for
the maximum mesh index in Section \ref{sec:dfal}, software defaults are used
throughout with the direction type set to ``{\tt OrthoMads n+1}'' 
\citep{AuIaLeDTr2012} and quadratic models~\citep{CoLed2011} disabled.
{\tt NOMAD} can handle constraints natively by using a progressive barrier
approach \citep{AuDe09a}, which we include as a representative comparator
from outside the APM framework.

{\bf Model-based:} We used the same code employed in \citet{kannan:wild:2012}.
A maximum of 50 blackbox evaluations were allotted to solving the subproblem
(\ref{eq:auglagsp}), with early termination being declared if the norm of the
gradient of the quadratically approximated AL was below $10^{-2}$; for the toy
problem this model gradient condition determined inner-loop termination, and
for the motivating hydrology problem in Section~\ref{sec:hydro} the budget was
The initial trust-region radius was taken to be $\Delta^0 = 0.2$ for the toy 
problem and $\Delta^0 = 10,000$ for the hydrology problem. 
In order to remain consistent with \citet{kannan:wild:2012}, a maximum of 5 
outer iterations (see Algorithm~\ref{alg:baseal}) were performed. If there 
remained function/constraint evaluations in the overall budget, the method was 
rerun from a random starting point (without incorporating any of the history of 
previous run(s)).

{\bf A note on relaxing convergence:} AL-based methods from the mathematical
programming literature tend to focus just outside of active constraints, so
examining only strictly feasible points may not lead to a fair comparison.
Therefore, we follow convention in constrained optimization and tolerate a
small degree of constraint violation when summarizing results for our
classical comparators: we consider a point $x^{(j)}$ to be effectively valid
if $\|\max(0,c(x^{(j)}))\|_{\infty}\leq 10^{-3}$.  Similar concessions are not
required for our EI-based methods, or other comparators.  Only strictly valid
results with $\|\max(0,c(x^{(j)}))\|_{\infty}\leq 0$ are reported for those
cases.

\subsubsection*{Other comparators}

To broaden the exercises we consider two further comparators: SA, as
representative of the stochastic optimization literature; asymmetric
entropy boundary exploration method of \citet{lind:lee:2015} as an alternative
class of methods discussed further in Section \ref{sec:discuss}.

For SA we use the \verb!method="SANN"! option in the {\tt optim} function
for {\sf R}, which is modeled after the \cite{belisle:1992} base specification.  Our
usage leverages default settings throughout.  To ensure that the default
settings, particularly for the random walk proposal mechanism, are appropriate
for our examples, we pre-scale the input bounding box $\mathcal{B}$ to lie in
the unit cube.  In our comparisons, we do not penalize SA  by
counting proposals lying outside $\mathcal{B}$ as blackbox evaluations against
the total budget. To accommodate blackbox constraints we deploy an APM
approach and sum a suitably scaled objective and (absolute value of)
constraint(s) so that the relative weightings in the composite are about equal
for each component.  This is meant to show SA in an idealized setting, as
generally such scales are unknown in advance.

The method of \citet{lind:lee:2015} uses a GP surrogate for the objective
function and a classification GP for the constraint, extending
\citet{gramacy:lee:2011}. However, rather than mixing the processes to form an
integrated improvement objective, it hybridizes entropy search
\citep{gramacy:polson:2011} to focus near the classification boundary (of the
valid region)---with ordinary EI---to bias the search away from the boundary and 
into
the interior of the valid region.  The search is started with a small
space-filling design (10 points for the toy problem, 100 for the hydrology
problem), and then additional points are chosen by maximizing the product of
expected improvement and the fifth power of asymmetric entropy with a mode of
$2/3$, following recommendations from \citeauthor{lind:lee:2015}.

\end{document}